\documentclass[twocolumn,showpacs,preprintnumbers,amsmath,amssymb,aps]{revtex4}
\usepackage{graphicx}
\usepackage{epsfig}
\usepackage{pict2e}
\def\lsim{\raise0.3ex\hbox{$\;<$\kern-0.75em\raise-1.1ex\hbox{$\sim\;$}}}
\def\gsim{\raise0.3ex\hbox{$\;>$\kern-0.75em\raise-1.1ex\hbox{$\sim\;$}}}
\newcommand{\be}{\begin{eqnarray}}
\newcommand{\ee}{\end{eqnarray}}

\newcommand{\n}{\nonumber\\}
\newcommand{\nn}{\tilde{\nu}}
\newcommand{\dd}{\displaystyle}
\def\bea{\begin{eqnarray}}
\def\eea{\end{eqnarray}}

\begin{document}
\title{Higgs Mass Corrections in the SUSY B-L Model with Inverse Seesaw}
\author{A. Elsayed$^{1,2}$, S. Khalil$^{1,3}$, and S. Moretti$^{4,5}$  }
\vspace*{0.2cm}
\affiliation{$^1$Center for Theoretical Physics at the
British University in Egypt, Sherouk City, Cairo 11837, Egypt.\\
$^2$Department of Mathematics, Faculty of Science, Cairo
University, Giza, 12613, Egypt.\\
$^3$Department of Mathematics, Faculty of
Science,  Ain Shams University, Cairo, 11566, Egypt.\\
$^4$School of Physics and Astronomy, University of Southampton,
Highfield, Southampton SO17 1BJ, UK.\\
$^5$Particle Physics Department, Rutherford Appleton Laboratory,
Chilton, Didcot, Oxon OX11 0QX, UK.}
%
%

\begin{abstract}
\noindent In the context of the Supersymmetric (SUSY) $B-L$
(Baryon minus Lepton number) model with an inverse seesaw
mechanism, we calculate the one-loop radiative corrections due to
right-handed (s)neutrinos to the mass of the lightest Higgs boson
when the latter is Standard Model (SM)-like. We show that such
effects can be as large as ${\cal O}(100)$ GeV, thereby giving an
absolute upper limit on such a mass around 200 GeV. The importance
of this result from a phenomenological point of view is twofold.
On the one hand, this enhancement greatly reconciles theory and
experiment, by alleviating the so-called `little hierarchy
problem' of the minimal SUSY realisation, whereby the current
experimental limit on the SM-like Higgs mass is very near its
absolute upper limit predicted theoretically, of 130 GeV. On the
other hand, a SM-like Higgs boson with mass below 200 GeV is still
well within the reach of the Large Hadron Collider (LHC), so that
the SUSY realisation discussed here is just as testable as the
minimal version.

\end{abstract}

\maketitle
%

The Higgs boson is the last missing particle in the SM. Higgs
boson discovery at the LHC is, therefore, crucial for its validity
as a low energy approximation of a new physics scenario valid to
high energy scales. A possibility for the latter emerges in SUSY
theories, wherein the Higgs mechanism is retained for mass
generation and multiple Higgs bosons appear in order  to cancel
anomalies. In addition, the stabilization of the Higgs mass
against loop corrections (gauge hierarchy problem) is possibly the
strongest motivation for a SUSY theory of nature. Hence, Higgs
boson discovery at the LHC is also crucial for SUSY as a whole. A
consequence of a SUSY Higgs sector is the existence of a stringent
upper bound on the mass of the lightest SUSY Higgs boson, $h$,
when the latter is SM-like. In the Minimal Supersymmetric Standard
Model (MSSM), this value is $m_h \lsim 130$ GeV. Therefore,
non-observing at the LHC a SM-like Higgs boson lighter than $130$
GeV would rule out the MSSM.

In detail, in the MSSM, the mass of the
lightest Higgs state can be approximated, at the one-loop level, as \cite{Ellis} %
\be%
m_h^2 \leq M_Z^2 + \frac{3 g^2}{16\pi^2 M_W^2}
\frac{m_t^4}{\sin^2\beta} \log\left(\frac{m_{\tilde{t}_1}^2
m_{\tilde{t}_2}^2}{m_t^4}\right),%
\label{deltat}
\ee%
where $g$ is the $SU(2)$ gauge coupling. $m_{\tilde{t}_{1,2}}$ are
the two stop physical masses. The ratio of the Electro-Weak (EW)
Vacuum Expectation Values (VEVs) is given by $\tan \beta
=v_2/v_1$. Note that the factor 3 in the above top-stop
correction is due to color. If one assumes that the
stop masses are of order TeV, then the one-loop effect leads to a
correction of order ${\cal O}(100)$ GeV, which implies
that %
\be%
m_h^{\rm MSSM} \lsim \sqrt{(90~{\rm GeV})^2 + (100~{\rm GeV})^2} \simeq 135~
{\rm GeV}.%
\ee%
It is worth mentioning that the two-loop corrections reduce this
upper bound by a few GeVs, to the aforementioned 130 GeV or so
value \cite{Heinemeyer:1999be}.

Experimental evidence now exists for physics beyond the SM, in the form of neutrino
oscillations, which imply neutrino masses \cite{Wendell:2010md}. In turn, the latter
imply new physics beyond not only the SM, but also the MSSM. Right-handed neutrino
superfields are usually introduced in order to implement the
seesaw mechanism, which provides an elegant solution for the
smallness of the left-handed neutrino masses. Right-handed neutrinos, which are heavy, can
naturally be implemented in the SUSY $B-L$ extension of the SM
(hereafter, the `SUSY $B-L$ model' for short), which is based on the gauge group $SU(3)_C \times
SU(2)_L \times U(1)_Y \times U(1)_{B-L}$. In this scenario, the scale
of $B-L$ symmetry breaking is related to the soft SUSY breaking
scale \cite{Khalil:2007dr}. Thus, the right-handed neutrino masses
are naturally of order TeV and the Dirac neutrino masses
must be less than $10^{-4}$ GeV ({\it i.e.}, they are of order the
electron mass) \cite{Khalil:2006yi}. Nevertheless, due to the
smallness of Dirac neutrino Yukawa couplings, the right-handed
neutrino sector has very suppressed interactions with the SM
particles. Therefore, the predictions of such a SUSY $B-L$ model
({\it i.e.}, with standard seesaw mechanism) remain close to
the MSSM ones. In particular, the discussed MSSM prediction for the lightest Higgs
boson mass upper bound remains intact. Same conclusion is obtained in
the context of the minimal Supersymmetric seesaw model, where the
right-handed neutrino masses are of order $10^{13}$ GeV
\cite{Heinemeyer:2010eg}.

The SUSY $B-L$ model with inverse seesaw, where Dirac neutrino
Yukawa couplings are of order 1, has recently been considered
\cite{Khalil:2010zz}. The superpotential of the leptonic sector
associated to this model is given by
{\small \begin{equation}%
W\!=\!Y_ELH_1E^c\!+\!Y_{\nu}LH_2N^c\!+\!Y_S N^c\chi_1S_2\!+\!\mu
H_1H_2\!+\!\mu'\chi_1\chi_2,%
\end{equation}}%
where $\chi_{1,2}$ are SM singlet superfields with $B-L$ charges
$+1$ and $-1$, respectively. Therefore, $U(1)_{B-L}$  is
spontaneously broken by the VEV of the scalar components of these
superfields. $N_i$ are three SM singlet chiral superfields with
$B- L$ charge $= -1$, introduced to cancel the $U(1)_{B-L}$
anomaly. The fermion components of $N_i$ account for right-handed
neutrinos. Finally, chiral SM singlet superfields $S_{1,2}$ with
$B - L$ charge $= +2,-2$ are considered to implement the inverse
seesaw mechanism. Note that a $\mathbb{Z}_2$ symmetry is assumed
in order to prevent the interactions between the field $S_1$ and
any other field.

After $B-L$ and EW symmetry breaking, the neutrino Yukawa
interaction terms lead to the following expression:%
\be%
{\cal L}_m^{\nu} = m_D \bar{\nu}_L N^c + M_N \bar{N}^c S_2,%
\ee%
where $m_D=Y_{\nu} v\sin\beta$ and $ M_N = Y_{S} v' \sin\theta$.
Light neutrino masses are related to a small mass term $\mu_S
\bar{S}_2^cS_2$, with $\mu_S\lsim{\cal O}(1)$ KeV, which can
emerge at the $B-L$ scale from a non-renormalizable term in the
superpotential, $\frac{\chi_1^4 S_2^2}{M_I^3}$, with $M_I$ an
intermediate scale of order ${\cal O}(10^7)$ GeV. Note that the
non-renormalizable scale $M_I$ can be related to a more
fundamental scale and couplings of the $S_2$ and $\chi$ fields
with integrated out fields and a suppression factor. In this
case one can write for instance $1/M_I^3 \sim \lambda^4/M_*^3$,
therefore, if $\lambda \sim {\cal O}(0.01)$, then $M_* $ is of
order $10^{12}$ GeV. Thus, the Lagrangian of neutrino masses, in the flavor basis, is given by: %
\be%
{\cal L}_m^{\nu} = m_D \bar{\nu}_L N^c   + M_N \bar{N}^c S_2 + \mu_{S} \bar{S_2^c}S_2.%
\ee%
In the basis $\{\nu_L, N^c, S_2\}$, the $3\times 3$ neutrino mass
matrix of one generation takes the form:%
\be\dd
{\cal M}_{\nu}=\left(%
\begin{array}{ccc}
  0 & m_D & 0\\
  m_D & 0 & M_N \\
  0 & M_N & \mu_{S}\\
\end{array}%
\right). %
\ee%
Therefore, the following light and heavy neutrino masses
\vspace{0.5cm}
are given by%
{\small \be \dd%
m_{\nu_{\ell}}&=& \frac{m_D^2\mu_{S}}{M_N^2+m_D^2},\\
m_{\nu_{H,H'}}&=&\pm \sqrt{M_N^2+m_D^2}+\frac{1}{2}\frac{M_N^2
\mu_{S}}{M_N^2+m_D^2}.%
\ee} %
In the limit of neglecting $\mu_S$, the neutrino masses are
{\fontsize{10}{3}\selectfont approximated as
\be
m_{\nu_{\ell}}^2 \simeq  0,\qquad\qquad~ m_{\nu_{H,H'}}^2 \simeq m_D^2+M_N^2.%
\label{m-nuaprox}%
\ee
}
\noindent
Further, in this type of model, the heavy neutrinos may have large
couplings to SM particles, leading to very interesting
phenomenological implications \cite{Abdallah:2011ew}.

The sneutrino mass matrix is obtained from the sneutrino scalar
potential, which is given by%
\be \dd
V_{\tiny\textnormal{scalar}}=V_F+V_D+V_{\tiny\textnormal{soft}},
\ee%
where $V_F$ is defined as usual as $\vert \partial W/\partial \phi
\vert^2$ and
{\small
\begin{eqnarray}\dd V_D \!=\!\frac{M_Z^2}{2}\cos
2\beta\nn_L^{\ast}\nn_L\!+\!M_{Z'}^2\cos
2\theta\Big(\!\frac{\nn_L^{\ast}\nn_L\!-\!\nn_R^{\ast}\nn_R}{2}\!+\!\tilde{S}_2^{\ast}\tilde{S}_2\Big),~
\end{eqnarray}}%
where $M_{Z'}$ is the mass of the $B-L$ neutral gauge boson~$Z'$,
given by $ M^2_{Z'}=4g''^2 v^{\prime 2}$. From the LEP II
experimental limits, one finds $M_{Z'}/g''>6\ {\rm TeV}$
\cite{Carena:2004xs}. Finally, $V_{\tiny\textnormal{soft}}$ is
defined as
{\small \be \dd
&&\!\!\!\!\!\!\!\!\!\!\! V_{\tiny\textnormal{soft}}=m_0^{2}
\!\sum_{\phi}\! \vert \phi \vert^2 \!\! + \!
\frac{1}{2}M_{1/2}\!\sum_i
{\widetilde{\lambda}}_i{\widetilde{\lambda}}_i + \Big[
A_0\Big(Y_{\nu} {\widetilde{N}}^{c} {\widetilde{L}}H_{2}
+ Y_{e}{\widetilde{E}}^{c}{\widetilde{L}}H_{1}\n%
&&\!\!\!\!\!\!\!\qquad+Y_{S}{\widetilde{N}}^{c}{\widetilde S_{2}}
\chi_1\!\Big)\!\!+\!\! B_0 \Big(\mu H_1 H_2 \!+\! \mu' \chi_1
\chi_2\Big) \!+\! h.c.\Big].~~ \label{soft}
\ee}
\noindent
Here, the sum in the first term runs over $\phi=H_1, H_2, \chi_1,
\chi_2, \tilde L, \tilde E^c,\tilde N^c,\tilde S_1, \tilde S_2$
and the sum in the second term runs over the gauginos: $\lambda_i=
{\widetilde{B}}, {\widetilde{W}}^a, {\widetilde{g}}^a,
{\widetilde{Z'}}$.

In general, one finds that the sneutrino mass matrix, for one
generation, can be written as a $3\times 3$ matrix, with entries
multiplied by the identity $2\times 2$ matrix
\cite{Khalil:2011tb}, {\it i.e.}, with one generation, one obtains
two left-handed sneutrinos $\tilde{\nu}_{L_{1,2}}$ and four
right-handed sneutrinos $\tilde{\nu}_{H_{3,4,5,6}}$:%
{\small \begin{widetext}
\begin{eqnarray}\label{snu-mass-matrix}
{\cal M}_{\nn}^2=\left(\begin{array}{ccc}
m_{\tilde{L}}^2+m_D^2+\frac{M_Z^2\cos 2\beta+M_{Z'}^2\cos 2\theta}{2} & m_D(A_{\nu}+\mu\cot\beta) & m_D M_N\\
m_D(A_{\nu}+\mu\cot\beta) & m_{\tilde{N}}^2+m_D^2+M_N^2-\frac{M_{Z'}^2}{2}\cos 2\theta & M_N(A_S+\mu'\cot\theta)\\
m_D M_N & M_N(A_S+\mu'\cot\theta) &
m_{\tilde{S}}^2+M_N^2+M_{Z'}^2\cos 2\theta
\end{array}\right).%
\end{eqnarray}
\end{widetext}}
If one chooses the $A$-terms such that elements $12$ and $23$
vanish, then the sneutrino masses can be written as%
{\small \be m_1^2&=&d,\n
m_2^2&=&\frac{1}{2}\left[(a+f)+\sqrt{(a-f)^2+4c^2}\right],\n
m_3^2&=&\frac{1}{2}\left[(a+f)-\sqrt{(a-f)^2+4c^2}\right], \ee
\be %
\textnormal{where \ }a&=& m_{\tilde{L}}^2+m_D^2+\frac{1}{2}(M_Z^2\cos
2\beta+M_{Z'}^2\cos 2\theta),\n %
c &=& m_D M_N,\n%
d &= & m_{\tilde{N}}^2+m_D^2+M_N^2-\frac{1}{2}M_{Z'}^2\cos
2\theta,
\n %
f &=& m_{\tilde{S}}^2+M_N^2+M_{Z'}^2\cos 2\theta.%
\ee}
If one assumes that $m_{\tilde{L}}=m_{\tilde{N}}=m_{\tilde{S}}=
\tilde{m}$ and neglects the $D$-term, then the sneutrino masses can be written as%
\be %
m_{\tilde{\nu}_{L_{1,2}}}^2=\tilde{m}^2,~~~~~
m_{\tilde{\nu}_{H_{3,4,5,6}}}^2 = m_D^2+M_N^2 + \tilde{m}^2. %
\label{m-snuaprox}%
\ee%

It is important to note that, unlike the squark sector, where only
the third generation (stops) has a large Yukawa coupling with the
Higgs boson, hence giving the relevant contribution to the Higgs mass
correction, all three generations of the (s)neutrino sector may
lead to important effects since the neutrino Yukawa couplings are
generally not hierarchical. Also, due to the large mixing between
the right-handed neutrinos $N_i$ and $S_{2_j}$, all the
right-handed sneutrinos $\tilde{\nu}_H$ are coupled to the Higgs
boson $H_2$, hence they can give significant contribution to the
Higgs mass correction. In this respect, it is useful to note that
the stop effect is due to the running of 6 degrees of freedom (3
colors times 2 stop eigenvalues) in the Higgs mass loop
corrections, while in case of right-handed sneutrinos we have, in
general, 12 degrees of freedom (3 generations times 4
eigenvalues).

As example of a generic $3\times 3$ neutrino Yukawa coupling,
$Y_\nu$, we consider
$Y_\nu = m_D/v_2$, with the Dirac neutrino mass matrix $m_D$ \cite{Abdallah:2011ew}%
{\small \begin{equation}
m_D=U_{\small\textnormal{MNS}}\,\sqrt{m_{\nu_{\ell}}^{\small\textnormal{phys}}}\,R\,\sqrt{\mu_S^{-1}}\,M_N,
\end{equation}}
\noindent
where $R$ is an arbitrary orthogonal matrix and
$U_{\small\textnormal{MNS}}$ is the light neutrino mixing matrix.
If we assume that {\small $R=I_{_{3\times 3}}$} and {\small
$\sqrt{m_{\nu_{\ell}}^{\small\textnormal{phys}}/\mu_S} \sim {\cal
O}(0.1)$}, then we find $Y_\nu \simeq U_{\small\textnormal{MNS}}$.
Note that here we assume a hierarchical $\mu_s$ in order to
account for a possible hierarchy between light neutrino masses.
For simplicity, we also assume universal Majorana neutrino masses,
{\small $M_{N}={\rm diag}\{M,M,M\}$}. In this case, one can easily
verify that the 12 right-handed sneutrinos have a very similar
mass coupled to with the Higgs boson~$H_2$ with order one Yukawa.

The one-loop radiative correction to the effective potential is
given by the
relation %
{\small \be%
\Delta V=\frac{1}{64\pi^2}\textnormal{
STr}\left[M^4\left(\log\frac{M^2}{Q^2}-\frac{3}{2}\right)\right],%
\ee}%
where $M^2$ is the field dependent generalized mass matrix
and $Q$ is the renormalization scale, to be fixed from minimization
conditions. The supertrace is defined as~follow:
\begin{center} ${\small
\textnormal{STr}f(M^2)=\sum_i(-1)^{2J_i}(2J_i+1)f(m_i^2).}$%
\end{center}
Here $m^2_i$ is a field dependent squared mass eigenvalue of a
particle with spin $J_i$. Therefore $\Delta V$, due to one
generation of neutrinos and sneutrinos, is given by %
{\small \be%
\!\!\Delta V_{\nu,\nn}\!=\!\frac{1}{64\pi^2}\Big[\!\!\sum_{i=1}^6
\!\!m_{\nn_i}^4\!\Big(\!\log\frac{m^2_{\nn_i}}{Q^2}\!-\!\frac{3}{2}\!\Big)
\!-\! 2 \!\!\sum_{i=1}^3
\!\!m_{\nu_i}^4\!\Big(\!\log\frac{m^2_{\nu_i}}{Q^2}\!-\!\frac{3}{2}\!\Big)\!\Big].~%
\ee}%
The first sum runs over the sneutrino mass eigenvalues, while the
second sum runs over the neutrino masses (with vanishing
$m_{\nu_1}$). In case of three generations, these sums should be
from 1 to 18 and from 1 to 9, respectively. In case of our above
example, where $Y_\nu \sim U_{\small\textnormal{MNS}}$, one finds
that the total $\Delta V$ is given by three times the value of $\Delta
V$ for one generation. This factor then compensates the color factor
of (s)top contributions.

The one-loop minimization conditions are given by $\frac{\partial
(V+\Delta V_{\nu\tilde{\nu}})}{\partial H_i}=0$, where $i=1,2$ and
$V= V_0 + \Delta V_{\small\textnormal{MSSM}}$. In order to retain
the minimization conditions as  $\frac{\partial V}{\partial
H_i}=0$, either we choose a renormalization scale $\hat{Q}$ such
that  $\frac{\partial \Delta V_{\nu\tilde{\nu}}}{\partial
H_i}\vert_{\hat{Q}}=0$, then we must evaluate the Higgs mass
correction at this scale, or we define the mass parameters $m_i^2$
in the potential $V(H_1,H_2)$ as%
$$m_i^2 = m_i^2\vert_{\small\textnormal{tree}} - \frac{1}{2 H_i}
\frac{\partial \Delta V_{\nu\tilde{\nu}}}{\partial
H_i}\vert_{H_i=v_i}.$$
\noindent
In this case, the genuine $B-L$ correction to the CP-even Higgs
mass matrix, due to the (s)neutrinos, at any renormalization scale
$Q$, is given by%
\be%
\Delta M_{ij}^2 = \frac{1}{2}\left[\frac{\partial^2 (\Delta V)_{\nu,\nn}}{\partial
H_i \partial H_j} - \frac{\delta_{ij}}{2 H_i} \frac{\partial
\Delta
V_{\nu\tilde{\nu}}}{\partial H_i}\right]_{H_i=v_i}.%
\ee
As known, the complete effective potential is scale independent.
However, the effective potential at one-loop level contains
implicit dependence on the scale. The $Q$-dependence is
approximately cancelled by neglecting the $D$-term and imposing
the minimization conditions as explained above.

From the (s)neutrino masses, given in
Eqs.~(\ref{m-nuaprox}) and~(\ref{m-snuaprox}), one can easily show that %
\bea%
\delta M_{11} &=& \delta M_{12} = \delta M_{21}=0,\\
\delta M_{22} &=& \frac{1}{16\pi^2}\left[\left(\frac{\partial
m_{\tilde{\nu}_H}^2}{\partial
v_2}\right)^2\log\frac{m_{\tilde{\nu}_H}^2}{\hat{Q}^2}-\left(\frac{\partial
m_{\nu_H}^2}{\partial v_2}\right)^2\log\frac{m^2_{\nu_H}}{\hat{Q}^2}\right]\n &
=&\frac{m_D^4}{4\pi^2
v_2^2}\,\log\frac{m_{\tilde{\nu}_H}^2}{m_{\nu_H}^2}. %
\label{deltanu} %
\eea%
Therefore the complete one-loop matrix of squared CP-even Higgs
masses will be given by $M_{\small\rm{tree}}+~\Delta M$, with~
$\Delta M
={\small \left(\begin{array}{cc} 0 & 0\\
0 & \delta_{t}^2 + \delta_{\nu}^2\end{array}\right)},$ %
where $\delta_t^2$ refers to the (s)top contribution presented in
Eq. (\ref{deltat}) and $\delta_\nu^2$ is the (s)neutrino correction
given in Eq. (\ref{deltanu}). In this case,  the
lightest Higgs boson mass is given by%
\begin{widetext}
\bea
m_h^2 =\frac{1}{2}(M_A^2+M_Z^2+\delta_t^2+\delta_\nu^2)\left[1-\sqrt{1-4\frac{M_Z^2 M_A^2\cos^22\beta+(\delta_t^2+\delta_\nu^2)(M_A^2\sin^2\beta+M_Z^2\cos^2\beta)}{(M_A^2+M_Z^2+\delta_t^2+\delta_\nu^2)^2}}\right]
\eea%
\end{widetext}
For $M_A\gg M_Z$ and $\cos2\beta\simeq 1$, one finds that
\begin{equation}
m_h^2 \simeq M_Z^2+\delta_t^2+\delta_\nu^2.
\end{equation}
If $\tilde{m}\simeq{\cal O}(1)$ TeV, $Y_\nu\simeq{\cal O}(1)$ and $M_N\simeq{\cal O}(500)$ GeV, one finds that $\delta_\nu^2\simeq{\cal O}(100)^2$, thus the Higgs mass will be of order $\sqrt{(90)^2+{\cal O}(100)^2+{\cal O}(100)^2}\simeq 170$ GeV.

\begin{figure}[t]
\includegraphics[scale=0.7]{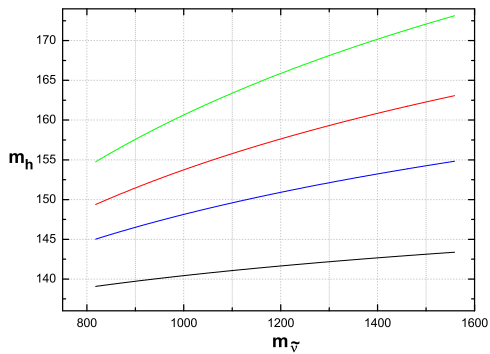}
\vspace{-0.5cm}\caption{\fontsize{8.5}{8.5}\selectfont Lightest
Higgs boson mass versus the right-handed sneutrino mass for
$M_N=500$ GeV,
$Y_\nu =0.8, 1, 1.1, 1.2$ (for curves from bottom to up respectively).} %
\label{mh-mnu}
\end{figure}
%

In Fig. \ref{mh-mnu} we show the Higgs mass, $m_h$, as a function
of the sneutrino mass, $m_{\tilde{\nu}}$, for $M_N=400$ GeV, and
$Y_\nu=0.8,1,1.1,1.2$. As shown from this figure, the neutrino
Yukawa coupling, $Y_\nu$, plays a crucial rule in inhancing the
lightest Higgs mass, $m_h$. Indeed, $Y_\nu$ must be of order one
(as natural in inverse seesaw) to be able to generate a sizable
(s)neutrino correction to $m_h$.

\begin{figure}[t]
\includegraphics[scale=0.7]{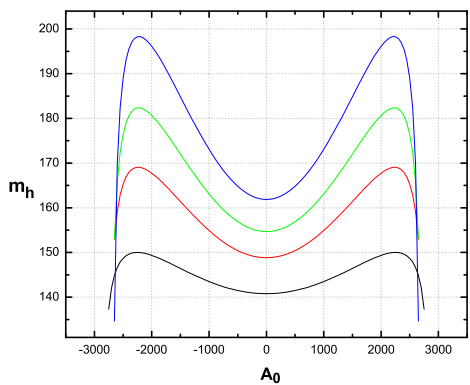}
\vspace{-0.5cm}\caption{\fontsize{8.5}{8.5}\selectfont Lightest
Higgs boson mass as a function of the triliniar coupling $A_N$ for
$M_N=400$ GeV,
$\tilde{m}=1$ TeV, and $Y_\nu =0.8, 1, 1.1, 1.2$ (for curves from bottom to up respectively).} %
\label{mh-An}
\end{figure}

Finally, we consider the impact of the trilinear couplings $A_N$
and $A_S$, which contribute to the off-diagonal elements of the
sneutrino mass matrix (\ref{snu-mass-matrix}). For sim\-pli\-ci\-ty, we
assume $A_N=A_S=A_0$. It turns out that, for a large value of
$A_0$,  $m_h$ can be enhanced by 20 GeV. In Fig. \ref{mh-An} we
display the dependence of $m_h$ on $A_0$ for
$Y_\nu=0.8,1,1.1,1.2$, $\tilde{m}=1$ TeV and $M_N=500$ GeV

%

In conclusion, we have calculated the one-loop radiative
corrections to the lightest SM-like Higgs boson mass, due to the
right-handed (s)neutrinos in a SUSY $B-L$ extension of the SM with
inverse seesaw mechanism. We have shown that the upper bound on
the Higgs mass can be enhanced to be around {200} GeV. This
enhancement alleviate a possible conflict between the experimental
limits from Higgs searches at the LHC and the absolute upper limit
predicted in MSSM theoretically, of 130 GeV. It is remarkable that
our result remains valid for any model beyond the MSSM with TeV
scale right-handed neutrinos (including Left-Right, Pati-Salam and
other models derived from $SO(10)$).


The work of A.E. and S.K. is partially supported by the Science
and Technology Development Fund (STDF) project 1855 and the ICTP
project 30. The work of S.M. is partially supported by the NExT
Institute. S.K. acknowledges an International Travel Grant from
the Royal Society (London, UK). A.E. thanks W. Abdallah for
fruitful discussions.


\end{document}